\documentclass[a4paper,UKenglish,cleveref, autoref]{lipics-v2019}
\usepackage{graphicx}
\usepackage[utf8]{inputenc}
\usepackage{amsmath}
\usepackage{algorithm}
\usepackage{algpseudocode}
\title{New Results and Bounds on Online Facility Assignment Problem}

\titlerunning{New Results and Bounds on Facility Assignment Problem}

\author{Saad Al Muttakee}{ Graph Drawing \& Visualization Lab , Department of Computer Science \& Engineering} {Bangladesh University of Engineering and Technology, Dhaka, Bangladesh}{}{}

\author{Abu Reyan Ahmed}{University of Arizona,  Tucson, Arizona, USA}{}{}{}

\author{Md. Saidur Rahman} {Graph Drawing \& Visualization Lab, Department of Computer Science \& Engineering}{Bangladesh University of Engineering and Technology, Dhaka, Bangladesh}{}{}

\authorrunning{Muttakee et al.}

\Copyright{Saad Al Muttakee, Abu Reyan Ahmed, Md. Saidur Rahman}

\ccsdesc[500]{Theory of computation~Online algorithms}

\keywords{Online problem, competitive ratio, facility assignment, grid graphs, voronoi diagram, cow path problem}

\category{}

\relatedversion{}

\supplement{}

\nolinenumbers 

\begin{document}

\maketitle

\begin{abstract}
Consider an online facility assignment problem where a set of facilities $F = \{ f_1, f_2, f_3, \cdots, f_{|F|} \}$ of equal capacity $l$ is situated on a metric space and customers arrive one by one in an online manner on that space. We assign a customer $c_i$ to a facility $f_j$ before a new customer $c_{i+1}$ arrives. The cost of this assignment is the 
distance between $c_i$ and $f_j$.
The objective of this problem is to minimize the sum of all assignment costs. Recently Ahmed et al. (TCS, 806, pp. 455-467, 2020) studied the problem where the facilities are situated on a line and computed competitive ratio of "Algorithm Greedy" which assigns the customer to the nearest available facility. They computed competitive ratio of algorithm named "Algorithm Optimal-Fill" which assigns the new customer considering optimal assignment of all previous customers. They also studied the problem where the facilities are situated on a connected unweighted graph.

In this paper we first consider that $F$ is situated on the vertices of a connected unweighted grid graph $G$ of size $r \times c$ and customers arrive one by one having positions on the vertices of $G$.
We show that Algorithm Greedy has competitive ratio $r \times c + r + c$ and Algorithm Optimal-Fill has competitive ratio $O(r \times c)$. We later show that the competitive ratio of Algorithm Optimal-Fill is $2|F|$ for any arbitrary graph. Our bound is tight and better than the previous result. We also consider the facilities are distributed arbitrarily on a plane and provide an algorithm for the scenario. We also provide an algorithm that has competitive ratio $(2n-1)$. Finally, we consider a straight line metric space and show that no algorithm for the online facility assignment problem has competitive ratio less than $9.001$.
\end{abstract}

\section{Introduction} \label{sec:one}
Let $F = \{ f_1, f_2, \cdots , f_{|F|} \}$ be a set of facilities, each with capacity $l$. The facilities are located on a metric space $M$. An input sequence $I = \{ c_1, c_2, \cdots , c_n \}$ is a set of $n$ customers who arrive one at a time in an online manner, with $c_i$ corresponding to the location of customer $i$ on $M$. 
The algorithms will assign the customer before the next customer appears. The objective is to minimize the total cost of all assignments. We call this problem the online facility assignment problem. This problem arises naturally in different practical applications, such handling online orders for a restaurant with multiple locations, and handling network packets in network with multiple routers. Consider a high performance computing machine which utilizes a set of interconnected processors to compute a large amount of jobs scheduled by the users of an organization. A natural way to speed up the processing time is to use parallel computing: dividing a job into small jobs and assign them into the set of processors. This problem can be modeled using the online facility assignment problem and a good assignment of jobs to processors will reduce the communication time and increase the performance of the overall system. 

\subsection{Related Works}
Ahmed et al. \cite{online_facility_assignment_tcs} first introduced the problem and calculated the competitive ratio for facilities on a line graph and facilities on a connected unweighted graph.
The offline version of facility assignment problem can be modeled as transportation problem \cite{lawler76,hitchcock41,schrijver2002history,ford1956solving}. In this scenario all the location of facilities and customers are known beforehand. This can also be thought as a variant of \textit{facility location problem}.
The online facility assignment problem is related to the $k$-server problem proposed by Manasse et al. \cite{Manasse:1990:CAS:82747.82753}, which decides the scheduling pattern of a set of $k$ servers will process the request coming in an online pattern. The servers can move in the plane considering the limitations. There is a famous conjecture related to the $k$-server problem which has been proven for some special cases~\cite{Sleator:1985:AEL:2786.2793,Chrobak90newresults,Chrobak:1991:OOA:103123.103131}. In online facility assignment problem the facilities are static and customers arrive in an online manner. \par

A recently proposed facility location variant is the $r$-gathering problem. An $r$-gathering of a set of customers $C$ for a
set of facilities $F$ is an assignment of to open facilities $F$
such that at least $r$ customers are assigned to each open
facility. There is a cost for assigning a customer to a facility and the objective is to minimize the total assignment cost.
The $r$-gathering problem was independently introduced by Karger and Minkoff \cite{karger892329} and by Guha et al. \cite{guha2000hierarchical}(who called it load-balanced facility-location). \par
The facility assignment problem can be seen as a generalization of the matching problem\cite{schrijver2002history}. Here each facility has capacity $l \geq 1$. In the online matching problem the facilities correspond to the right side of a bipartite graph. 

The customers appear in an online manner as vertices on the left side of the graph and each customer must be assigned to a facility before the next customer appears. This problem was first independently introduced by Khuller et al. \cite{KHULLER1994255} and Kalyanasundaram et al. \cite{KALYANASUNDARAM1993478}. \par

The previous results inspired the \textit{Facility Assignment Problem}. Facility assignment problem is discussed in detail at \cite{online_facility_assignment_tcs}. The static facilities are set in different embedding and the customers arrive in online manner. Assigning any customer to a facility has a cost. The cost is measured by the distance between the customer and the facility. The ultimate goal is to minimize the total assignment cost of all the customers. The customers can appear on a plane or on a connected unweighted graph. The competitive ratios proven in \cite{online_facility_assignment_tcs} holds firm whether the set of customers are well-distributed or not. At first the authors considered  the case where both the facilities $|F|$ and the customers $|C|$ are on a straight line. The proposed Algorithm Greedy has competitive ratio $4|F|$. Introducing randomization in Algorithm Greedy leads to an improved performance of $\frac{9}{2}$ for a special class of input instances. The authors then described Algorithm Optimal-Fill and show it has competitive ratio $|F|$.
Then the authors assumed the facilities and the customers are located on the vertices of an unweighted graph $G =(V, E)$. Algorithm Greedy for this scenario has competitive ratio $2|E(G)|$ and Algorithm Optimal-Fill has competitive ratio $\frac{|E(G)||F|}{r}$, where $r$ is the radius of $G$. Finally, They also briefly discussed on the case where a customer leaves after receiving service at a facility.

\subsection{Our contributions}
In this paper we first consider the case where the facilities are on a grid graph and show that Algorithm Greedy on grid graph has the competitive ratio of $r \times c + r + c$. We also show that the competitive ratio of Algorithm Optimal-Fill on grid graph is $O(r \times c)$.
The competitive ratio of Algorithm Optimal-Fill on arbitrary connected unweighted graph is $2|F|$. Our bound is tight and better than the previous result $\frac{|E||G||F|}{r}$, where $E$ is the set of edge of graph $G$, $F$ is the set of facilities and $r$ is the radius of the graph. \par

We then turn to a more generalized form and assume the facilities are situated arbitrarily on a plane. We use Voronoi diagram to distribute the plane to existing facilities and assign the customers according to their capacity. The algorithm has a competitive ratio of $(2n-1)$ on the plane. \par

We then consider a straight line metric space. We show that no algorithm for the online facility assignment problem has competitive ratio less than $9.001$. To do this we establish that the cow path problem and the facility assignment on a line problem from \cite{online_facility_assignment_tcs} are bijectional. \par

The rest of this paper is organized as follows. In Section \ref{sec:preliminaries} we give some defintions and terminologies. In Section \ref{sec:two} we study facility assignment on a grid graph. In Section \ref{sec:three} we study facility assignment on a connected unweighted graph. In Section \ref{sec:four} we study facility assignment on a plane. Finally, in Section \ref{sec:five} we analyze the hardness result for facility assignment on a line using cow-path problem.

\section{Preliminaries} \label{sec:preliminaries}
\label{Preliminaries}
A graph $G=(V, E)$  consists of a finite set $V$ of vertices and a finite set $E$ of edges; each edge is an unordered pair of vertices. We often denote the set of vertices $G$ by $V(G)$ and the set of edges by $E(G)$. 
A $r \times c$-grid graph is a graph whose vertices correspond to the grid points of a $r \times c$-grid in the plane and edges correspond to the grid lines between two consecutive grid points.
We say $G$ is \textit{unweighted} if every edge of $G$ has equal weight. Let $u$ and $v$ be two vertices of $G$. If $G$ has a $u,v$-path, then the distance from $u$ to $v$ is the length of a shortest $u,v$-path, denoted by $d_G(u,v)$ or simply by $d(u, v)$. If $G$ has no $u, v$-path then $d(u, v) =\infty$. The \textit{eccentricity} of a vertex $u$ in $G$ is $max_{v \in V(G)} d(u, v)$ and denoted by $\epsilon(u)$. The \textit{radius} $r$ of $G$ is $min_{u \in V(G)} \epsilon(u)$ and the \textit{diameter} of $G$ is $max_{u \in V(G)} \epsilon(u)$. The \textit{center} of $G$ is the subgraph of $G$ induced by vertices of minimum eccentricity. 

In the online facility assignment problem, we are given a set of facilities $F=\{f_1,f_2,$ $\cdots,f_{|F|}\}$ of equal capacity $l$  in a metric space, and an input sequence of customers $I=\{c_1,c_2,\cdots,c_n\}$ which is a set of $n$ customers who arrive one at a time in an online manner, with $c_i$ corresponding to the location of customer $i$ in the given space. We say an input $I$ is \textit{well distributed} if there is at least one customer between any two adjacent facilities. The capacity of a facility is reduced by one when a customer is assigned to it. We denote the current capacity of facility $f_i$ by $capacity_i$. A facility $f_i$ is called \textit{free} if $capacity_i>0$.  Any algorithm ALG for this problem must assign a customer $c_i$ to a free facility $f_j$ before a new customer $c_{i+1}$ arrives. The cost of this assignment is the distance between $c_i$ and $f_j$, which is denoted by $distance(f_j,c_i)$.
We now define the \textit{cover area} of a facility situated on a line. Consider a facility $f_i$ with two adjacent free facilities $f_j$ and $f_k$. Let $p_1$ and $p_2$ be the mid-points of  $(f_i, f_j)$ and $(f_i,f_k)$ respectively. The cover area of $f_i$ is then the line segment $p_1$ to $p_2$. The total number of customers is, at most, $|F|l$ (where $l$ is the capacity of a facility) and each customer must be assigned to a facility. For any input sequence of customers $I$, Cost\_ALG($I$) is defined as the total cost of all assignments made by ALG. The objective is to minimize Cost\_ALG($I$).

We say an algorithm is \textit{optimal} if, for any input sequence of customers, the total cost of the assignment it provides is the minimum possible. We denote an optimal algorithm by OPT and the cost of that algorithm by $\text{Cost\_OPT}$. An online algorithm ALG is $c$-competitive if there is a constant $\alpha$ such that, for all finite input sequences $I$, \[\text{Cost\_ALG}(I) \le c.\text{Cost\_OPT}(I) + \alpha.\]
The factor $c$ is called the {\em competitive ratio} of ALG.
When the \textit{additive constant} $\alpha$ is less than or equal to zero (i.e., $\text{Cost\_ALG}(I) \le c.\text{Cost\_OPT}(I)$), we may say, for emphasis, that ALG is \textit{strictly} $c$-competitive. An algorithm is called \textit{competitive} if it attains a constant competitive ratio $c$. Although $c$ may be a function of the problem parameters, it must be independent of the input $I$. The infimum over the set of all values $c$ such that ALG is $c$-competitive is called \textit{the competitive ratio} of ALG and is denoted by $\mathcal{R}(\text{ALG})$. \par
 
 We can analyze the online algorithm in the context of a game between an online player and a malicious adversary. The online player runs the online algorithm on an input created by the adversary. The adversary, based on the knowledge of the online algorithm, constructs the worst possible input (i.e., one that maximizes the competitive ratio). The adversary strategy of designing an instance very costly for the target algorithm but, at the same time, inexpensive for the optimal output. \par

 \section{Facility Assignment on a Grid} \label{sec:two}
In this section we assume that the facilities $F=\{ F_1, F_2, F_3, \cdots, F_{|F|} \}$ are situated on a grid graph and the customers $C = \{ c_1, c_2, c_3, \cdots, c_n \}$ arrive on the vertices of the graph in an online manner. We assign each customer to a facility before the next customer arrives. In Section \ref{sec:three_one}
we compute the competitive ratio of Algorithm Greedy on a grid graph and in Section \ref{sec:three_Two} we compute that of Optimal-Fill on a grid graph.
 

\subsection{Greedy Approach} \label{sec:three_one}
Algorithm Greedy assigns the current customer to its nearest unassigned facility. This process continues until all the facilities are filled up. We now prove the following theorem.

\begin{theorem}
\label{theorem:greedy_grid}
Let $\mathcal{M}$ be a grid graph of size $r \times c$. Then $\mathcal{R}(\text{Algorithm}$ $ \text{Greedy}) < r \times c + r + c$.
\end{theorem}
\begin{proof}
The input sequence can be either well distributed or not well distributed. We consider these two cases separately. For both cases, assume now that the facilities have unit capacity. Later we will also deal with the case for capacity $l$, where $l > 1$.

We first consider that the input sequence is well distributed. 
We illustrate a worst case scenario in Figure~\ref{figure:greedy_grid_tight}. There is a facility on every vertex of the graph except $v_2$. The first customer appears on $v_2$. Without loss of generality we assume that Algorithm Greedy assigns this customer to the facility on $v_3$. The adversary places the next customer on $v_3$. Since, the facility on $v_3$ is already assigned for the customer on $v_2$, Algorithm Greedy assigns the customer on $v_3$ to the facility on $v_4$. The adversary continues this process. The last customer appears on $v_{15}$. Algorithm Greedy assigns that customer to the facility on $v_1$. Hence, the assignment cost of the last customer is equal to $(r+c)$. In the optimal assignment, the customer on $v_2$ is assigned to the facility on $v_1$ and the remaining customers are assigned to facilities situated on the same vertex. Hence, the optimal assignment cost is equal to one. Hence, $\mathcal{R}(\text{Algorithm}$ $ \text{Greedy}) < r \times c + r + c$.

Note that in the example of Figure~\ref{figure:greedy_grid_tight}, we assume that Algorithm Greedy always makes the worst case assignment. For example, the customer on $v_6$ has two nearest facility: the facilities on $v_1$ and $v_7$. Algorithm Greedy could assign the customer to the facility on $v_1$ too. However, we assumed that it assigns the customer to the facility on $v_7$ to illustrate a worst case scenario. In fact the assumption is not invalid since both facilities have the same distance. However, it is possible to generate a similar scenario where such an assumption is not necessary. For example, consider the scenario illustrated in Figure~\ref{figure:another_greedy_grid}. The analysis is not tight in this case. However it is easy to show that $\mathcal{R}(\text{Algorithm}$ $ \text{Greedy}) \leq O(r\times c)$.


\begin{figure}[h]
\begin{center}
\includegraphics[width=6cm]{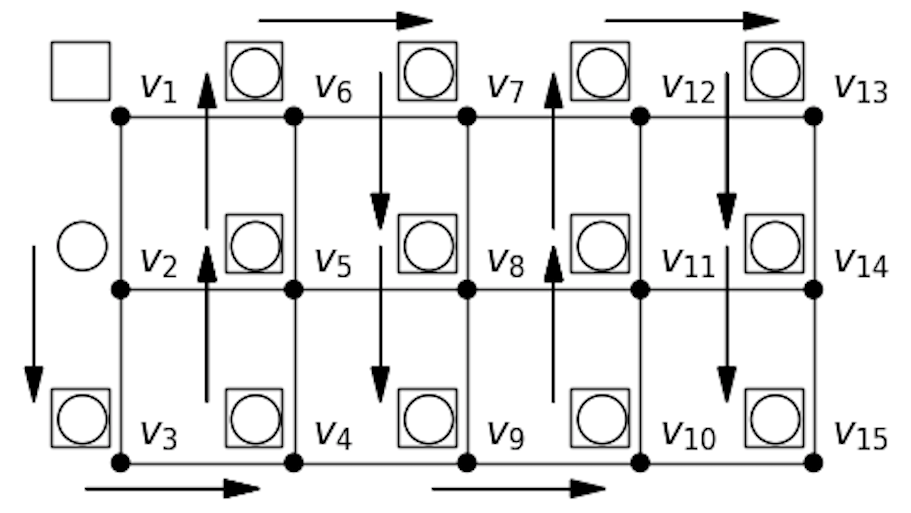}
\end{center}\caption{The worst case scenario of Algorithm Greedy}
\label{figure:greedy_grid_tight}
\end{figure}

In the second case, the input sequence is not well distributed. It is very simple to show that the ratio between $\text{Cost\_Algorithm\_Greedy}(I)$ and $\text{Cost\_OPT}(I)$ will not be greater than the ratio in the first case. The customers are concentrated in some small areas and the effect of different assignments is limited to only these spanning areas.  If the spanning areas are very small, no algorithm can save that much. In extreme case, when all customers are placed in the same location, all assignments are same.

\begin{figure}[h]
\begin{center}
\includegraphics[width=6cm]{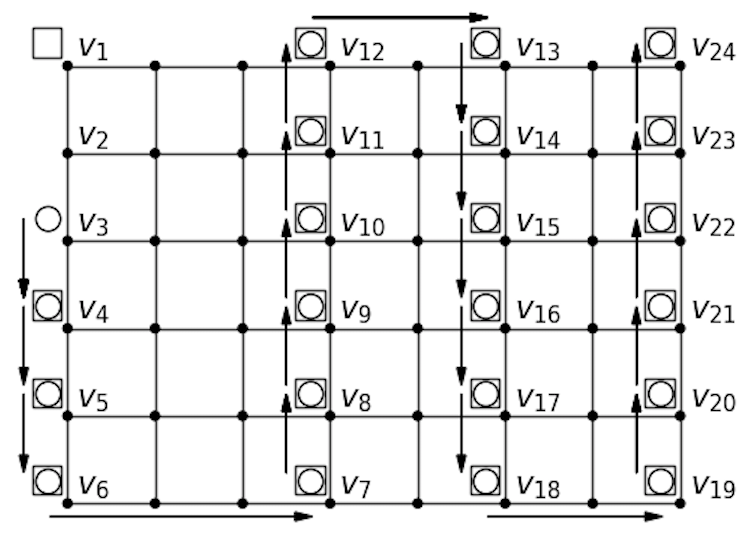}
\end{center}\caption{Another configuration where Algorithm Greedy shows poor performance}
\label{figure:another_greedy_grid}
\end{figure}

In the analysis above we assumed unit capacity; now  let each facility have capacity $l$, where $l > 1$. Suppose that there exists an input sequence of customers $I$ for which the ratio is greater than $r \times c + r + c$. We can partition $I$ into $I_1, I_2, \cdots, I_l$ in such a way that the following conditions hold:

\begin{itemize}
\item $I_i \cap I_j = \emptyset$ for $1 \leq i,j \leq l$ and $i \neq j$.
\item $I_1 \cup I_2 \cup \cdots \cup I_l = I$.
\item Exactly one customer from $I_i$ is assigned to a facility $f_j$ for $1 \leq i \leq l$ and $1 \leq j \leq |F|$.
\end{itemize}

Then there exists a set $I_{max} \in \{I_1 , I_2 , \cdots , I_l\}$ such that the ratio of the corresponding cost of Algorithm Greedy to the cost of OPT is greater than $r \times c + r + c$. If we take a set of facilities with unit capacities and place the customers of $I_{max}$ in the same order as they appear in $I$, the ratio would be greater than $r \times c + r + c$ which is a contradiction to the bound of unit capacity.

\end{proof}

\subsection{Optimal-Fill Approach} \label{sec:three_Two}

In the previous section we provided the analysis of Algorithm Greedy on a grid graph. We now provide the analysis of Algorithm Optimal-Fill~\cite{online_facility_assignment_tcs} on a grid graph. The idea behind this approach is that when a new customer $c_i$ arrives, it finds out the new facility $f_j$ that would be selected by an optimal assignment of all the customers $c_1, c_2, \cdots c_i$. Algorithm Optimal-Fill then assigns $c_i$ to $f_j$. In other words, consider the $i$th step of the algorithm, customer $c_i$ just appeared and the Algorithm Optimal-Fill has not assigned $c_i$ to any facility yet. Now, there are $i$ customers, $i-1$ of them are already assigned by the algorithm, let the set of facilities assigned for these customers is $F_{\text{Optimal-Fill}}$. Algorithm Optimal-Fill will find the optimal assignment of the $i$ customers. Suppose the set of facilities assigned by the optimal algorithm is $F_{\text{OPT}}$. For simplicity assume that each facility has unit capacity. Since, the optimal assignment considered $i$ customers and Algorithm Optimal-Fill assigned $i-1$ customers, $|F_{\text{OPT}}|-|F_{\text{Optimal-Fill}}|=1$. Algorithm Optimal-Fill selects the facility in $F_{\text{OPT}} \setminus F_{\text{Optimal-Fill}}$ and assign $c_i$ to that facility. The motivation behind this algorithm is if an algorithm follows the optimal assignment, then it can get rid of some traps that the adversary may set up like shown in Figure~\ref{figure:another_greedy_grid}. For example, when the adversary places the third customer on $v_5$, if we check the optimal assignment of three customers, then a natural action is to assign the customer on $v_5$ to the facility on $v_1$. Ahmed et al.~\cite{online_facility_assignment_tcs} provided the analysis of Algorithm Optimal-Fill when the metric space is a line. Here, we consider the problem on a grid which is a more general metric space.


\begin{lemma}
Let $G$ be a $n\times n$ grid graph where $n>1$. Let $c$ be the center of the graph. Let $V'$ be a set of vertices such that every vertex in $V'$ has equal distance from $c$. Then the size of $V'$ is at most $2n-2$.
\label{lemma:opt_fill_max_vertices}
\end{lemma}

\begin{lemma}
Let $G$ be a $n\times n$ grid graph. Let $c$ be the center of the graph. Let $V'$ be the set of vertices such that every vertex in $V'$ has distance equal to $r$ from $c$ where $r>0$. Then the size of $V'$ is at most $4r$.
\label{lemma:equal_distance_from_center}
\end{lemma}

\begin{lemma}
Let $G$ be a $n\times n$ grid graph. Let $c$ be the center of the graph. Let $V'$ be the set of vertices such that every vertex in $V'$ has distance equal to $r$ from $c$ where $r>0$. Every vertex $v$ in $V'$ has a facility situated on top of it. Then total assignment cost of Algorithm Optimal-Fill is $O(r^2)$.
\label{lemma:optimal_fill_total_cost}
\end{lemma}

The proofs of above lemmas are deferred to Appendix~\ref{apdx:opt_fill_max_vertices}, \ref{apdx:equal_distance_from_center}, and \ref{apdx:optimal_fill_total_cost}.

\begin{theorem}
\label{theorem:optimal_grid}
Let $\mathcal{M}$ be  a  grid of size $r \times c$. Then $\mathcal{R}(\text{Algorithm}$ $ \text{Optimal-Fill}) \leq O(r \times c)$.
\end{theorem}
\begin{proof}
We only assume the case where the input is well-distributed and every facility has unit capacity. The analysis of other cases is similar to Algorithm Greedy. In the worst case, there is a facility on every vertex except the center of the grid. The adversary places the first customer to the center vertex. Algorithm Optimal-Fill assigns that customer to the closest facility $f_1$. Then the adversary places the next customer exactly on $f_1$. However, Algorithm Optimal-Fill can not assign the new customer to $f_1$, since $f_1$ is already assigned to another customer. Suppose, Algorithm Optimal-Fill assigns the new customer to facility $f_2$. The adversary places the next customer on $f_2$ and continues this process. The situation is similar to Figure~\ref{figure:opt_fill_r_grid}.
For simplicity we assume that $r=c=n$. 
By Lemma~\ref{lemma:opt_fill_max_vertices} we know that there are $O(n)$ vertices in a set of vertices such that each has the same distance from the center. Since there are $O(n^2)$ vertices in total we can have $O(n)$ such set. By Lemma~\ref{lemma:optimal_fill_total_cost}, we know that the cost of total assignment cost corresponding to the set of vertices having an equal length from the center is $O(n^2)$. Since the graph is a grid, the radius of the graph is $O(n)$. Hence, in total the cost of Algorithm Optimal-Fill is $O(n^3)$. The optimal algorithm assigns the customer placed on the center vertex with cost $O(n)$ and the remaining customers do not have a significant assignment cost. Hence, the ratio of the cost of Algorithm Optimal-Fill to the optimal algorithm is $O(n^2)$.

We now consider the case where $r \neq c$. Without loss of generality, we assume that $r>c$. The equivalent result corresponding to Lemma~\ref{lemma:opt_fill_max_vertices} is the size of $V'$ is at most $O(c)$ where $V'$ is a set of vertices such that every vertex in $V'$ has equal distance from a center vertex. Consider the distance of a vertex in $V'$ from the center is $d$. If $d \leq c$, then it is trivial to show that the size of $V'$ is at most $O(c)$. Now, if $d>c$, then there are $O(d-c)$ rows for which there are no corresponding vertices in $V'$. Hence, the size of $V'$ is at most $O(c)$. Using a similar argument for the scenario $r=c$, we can show that the ratio of the cost of Algorithm Optimal-Fill to the optimal algorithm is $O(r\times c)$. 

\end{proof}

\section{Facility Assignment on Connected Unweighted Graphs}\label{sec:three}

In this section we analyze the competitive ratio of Algorithm Optimal-Fill on a connected unweighted graph. It is known from \cite{online_facility_assignment_tcs}  that the optimal fill approach has the competitive ratio of $\frac{|E||F|}{r}$, where $|E|$ is the number of edges, $|F|$ is the number of facilities and $r$ is the radius of the given graph. In this section we provide a tighter bound equal to $2|F|$. Before providing the main theorem, we first prove a useful lemma. 

\begin{lemma}
\label{lemma:optimal_lower_bound}
Assume that Algorithm Optimal-Fill assigned a customer on vertex $v_c$ to a facility on vertex $v_f$. Let the distance between $v_c$ and $v_f$ be equal to $x$. Let the closest facility from $v_c$ along the path from $v_c$ to $v_f$ be situated on vertex $v_{f'}$. Let the distance between $v_f$ to $v_{f'}$ be $x'$. Note that, $x\geq x'$. Then the cost of the optimal algorithm is at least $\frac{x'}{2}$.
\end{lemma}

The proof is deferred to Appendix~\ref{apdx:optimal_lower_bound}. 

\begin{theorem}
\label{theorem:optimal_graph}
Let $\mathcal{M}$ be a connected unweighted graph and let a set of facilities $F$ be placed on the vertices of $\mathcal{M}$. Then $\mathcal{R}(\text{Algorithm}$ $ \text{Optimal-Fill}) \leq 2|F|$.
\end{theorem}
\begin{proof}
We assume that the facilities have unit capacity since the analysis is similar to Theorem~\ref{theorem:greedy_grid} for capacity $l$, where $l > 1$. Two facilities $f_i$ and $f_j$ are \textit{adjacent} if there exists a path $P$ from $f_i$ to $f_j$ such that no other facilities are situated on $P$. 
Recall the definition of a well distributed input sequence: an input $I$ is \textit{well distributed} if there is at least one customer between any two adjacent facilities. 
We first prove the claim for an input $I$ which is well distributed. Then we show how to transform $I$ to $I'$ such that $I'$ is not well distributed and show that the competitive ratios of $I$ and $I'$ are the same.

We consider two cases; $\mathcal{M}$ has no cycle and $\mathcal{M}$ contains at least one cycle. If $\mathcal{M}$ does not have any cycle, there is only one path between two vertices. Consider the scenario shown in Figure~\ref{figure:opt_fill_path_cycle}, while assigning the last customer, Algorithm Optimal-Fill has to traverse the whole path. A square box represents a facility and the input customers are shown by their sequence numbers.

\begin{figure}[h]
\begin{center}
\includegraphics[width=7cm]{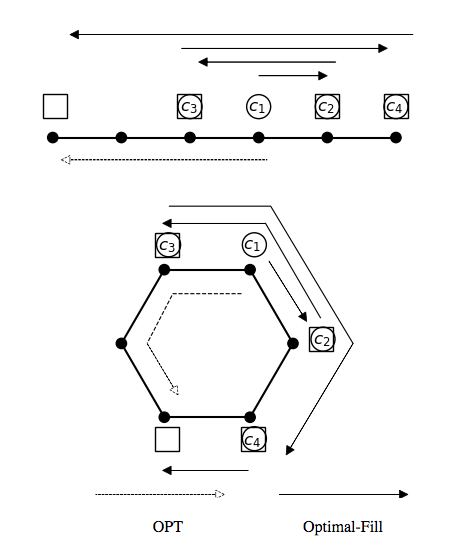}
\end{center}\caption{The configurations of Algorithm Optimal-Fill and OPT for a path and a cycle.}
\label{figure:opt_fill_path_cycle}
\end{figure}

If $\mathcal{M}$ contains a cycle, $\mathcal{R}(\text{Algorithm Optimal-Fill})$ does not increase. Consider a set of facilities $F$ are situated on a cycle. The scenario is almost same as before except one extra edge present in the graph due to the fact that it is a cycle. This extra edge may provide a shorter path to assign the last customer to a free facility as shown in Figure~\ref{figure:opt_fill_path_cycle}. Hence, in this case the ratio between $\text{Cost\_Optimal\_Fill}(I)$ and $\text{Cost\_OPT}(I)$ is less than the previous case.

\begin{figure}[H]
\begin{center}
\includegraphics[width=10.0cm]{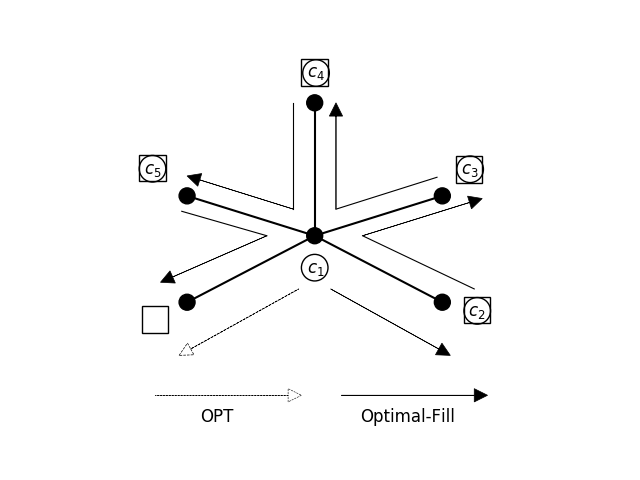}
\end{center}\caption{The worst case scenario of Algorithm Optimal-Fill and OPT for graphs.}
\label{figure:opt_fill_graph_worst_case}
\end{figure}

According to the above argument, the worst case scenario arise when $\mathcal{M}$ is a tree. The worst case scenario is shown in Figure~\ref{figure:opt_fill_graph_worst_case}. We will formally prove it later by providing a tight competitive ratio. Let $x$ be a vertex in the center of $\mathcal{M}$ which is not a facility. The first customer $c_1$ is placed on $x$. Algorithm Optimal-Fill pays a cost equal to the distance between $c_1$ and the closest free facility. 
For a customer $c_i$ ($i > 1$), let $f_{i-1}$ be the facility which has been assigned to $c_{i-1}$ by Algorithm Optimal-Fill when $c_{i-1}$ arrives. Let $f_i$ be the new facility used by OPT for assigning the customers $c_1, c_2, \cdots c_n$. 
Algorithm Optimal-Fill pays a cost equal to the distance between two facilities $f_{i-1}$ and $f_i$ for each customer $c_i$, except the first one (see Figure~\ref{figure:opt_fill_graph_worst_case}). The adversary pays a cost which is no more than radius only for the first customer. 
The assignment costs for each customer by Algorithm Optimal-Fill is no more than $2|F|r$, where $r$ is the radius of $\mathcal{M}$. The optimal algorithm only pays $r$ for the first customer. Hence, R(Algorithm Optimal-Fill) is at most $2|F|$.
Let $f$ be the closest facility from a leaf $b$ of $\mathcal{M}$. Consider a customer $c$ who appears between $b$ and $f$. The distance between $c$ and $f$ is $distance ( f , c )$. 
Both $\text{Cost\_Algorithm\_Optimal\_Fill} ( I )$ and $\text{Cost\_OPT} ( I )$ must pay the amount $distance ( f , c )$. The ratio of $\text{Cost\_Algorithm\_Optimal\_Fill} ( I )$ to $\text{Cost\_OPT} ( I )$ increases when $distance ( f , c )$ decreases. 
Hence, we can assume that there is a facility in every leaf of $\mathcal{M}$.
We now prove that for any input sequence, the ratio between the cost of Algorithm Optimal-Fill and the cost of OPT is no more than $2|F|$. 

An assignment cost of a customer can be at most $2r$. Hence, the total assignment cost of Algorithm Optimal-Fill is at most $2r|F|$. Then by pigeonhole principle at least one customer $c$ has cost at least $2r$. Let the customer is situated on vertex $v_c$. Assume that Algorithm Optimal-Fill assigned $c$ to a facility on vertex $v_f$. Note that, both $v_c$ and $v_f$ are two leaves since the distance between them is $2r$. This means there is also a facility on $v_c$ since we assumed that there is a facility on each leaf. By Lemma~\ref{lemma:optimal_lower_bound}, the cost of the optimal algorithm is at least $r$. Hence, the ratio of $\text{Cost\_Algorithm\_Optimal\_Fill} ( I )$ to $\text{Cost\_OPT} ( I )$ is no more than $2|F|$.

We now specifically present the gap between the analysis of Ahmed et al.~\cite{online_facility_assignment_tcs}. They have shown that the competitive ratio of Algorithm Optimal-Fill is $\frac{|E||F|}{r}$, where $E$ is the set of edges of graph $G$, $F$ is the set of facilities and $r$ is the radius of the graph. However, their analysis is not tight. In their argument, they mentioned that every edge will be counted at most $|F|$ times since there are $|F|$ customers in the unit capacity setting. Hence, the total assignment cost is $|E||G|$. This counting bound is not tight, since each assignment cost can be no more than $2r$. Hence, the total assignment cost is at most $2r|F|$.

Now suppose the input sequence $I$ is not well distributed. Let $\mathcal{M}'$ be the minimum subgraph of $\mathcal{M}$ so that all customers are situated 
on $\mathcal{M}'$. Consider the set of facilities $F'$ situated on $\mathcal{M}'$. In the worst case the customers assigned to those facilities by Algorithm Optimal-Fill incur total cost less than $|F'|$ and OPT incurs only $r'$, where $r'$ is the radius of $\mathcal{M}'$. If OPT incurs cost $x$ to assign a customer to a remaining facility, then Algorithm Greedy incurs at most $x+|E(\mathcal{M}')|$ cost to assign a customer to that facility. Hence, $\text{Cost\_Optimal\_Fill} ( I ) \le \text{Cost\_OPT} ( I ) - r' + | E(\mathcal{M}') | (|E(\mathcal{M}) | - | E(\mathcal{M}') | ) + 2|F'| $. It follows that if $|E(\mathcal{M}')|$ is small then Algorithm Optimal-Fill will perform similar to OPT. The larger the value of $|E(\mathcal{M}')|$ the more well distributed the input $I$ becomes. Hence $\mathcal{R}(\text{Algorithm}$ $ \text{Optimal-Fill}) \leq 2|F|$.
\end{proof}

\section{Facility assignment on a plane} \label{sec:four}

A Voronoi diagram is a partition of a plane into regions close to each of a given set of objects. In the simplest case, these objects are just finitely many points in the plane (called seeds, sites, or generators). For each seed there is a corresponding region consisting of all points of the plane closer to that seed than to any other. These regions are called Voronoi cells.

Let $P = \{ p_1, p_2, \cdots , p_n \} $ be a set of $n$ distinct points in the plane; these points are the sites.
We define the Voronoi diagram of $P$ as the subdivision of the plane
into $n$ cells, one for each site in $P$, with the property that a point $q$ lies in the
cell corresponding to a site $p_i$ if and only if $dist(q, p_i) < dist(q, p_j)$ for each
$p_j \in P$ with $j \leq i$. Here $dist(x,y)$ defines the euclidean distance between two points $x$ and $y$.

A weighted Voronoi diagram is a special case of a Voronoi diagram. The Voronoi cells in a weighted Voronoi diagram are defined in terms of a distance function. The distance function may specify the usual Euclidean distance, or may be some other, special distance function. Usually, the distance function is a function of the generator point's weights.
We divide the plane into $n$ cells and assign them to the nearest facilities. Each cell is assigned to a facility and we call each cell the \textit{area of influence} of the facility.
 
When a customer appears, we locate the area of influence where the customer is located. We assign the customer to the corresponding facility of that area of influence. The voronoi diagram is redrawn after adjusting the new weights(capacities) of the facilities. The process continues until all the facilities are filled up. We provide the algorithm in Appendix~\ref{apdx:Voronoi}.


\begin{theorem}
Let $F = \{ f_1, f_2, \cdots , f_n \}$ be a set of facilities situated on a $2D$ plane, then $\mathcal{R}(\text{Algorithm}$ $ \text{Capacity sensitive voronoi}) \leq 2n-1$.
\end{theorem}
\begin{proof}

For this proof we consider $n$ number of facilities inside a regular polygon with $n$ vertices. We can divide the the polygon into $n$ triangles. The facilities are situated in the centroid of the triangles. 
At first we assume each facility has unit capacity and they are well-distributed throughout the plane. Once a customer is assigned, the customer can not leave or interchange position with any other customer. Suppose the length of the side of the outermost square is $d$.
Let \textbf{cost(ALG)} be the measured cost of our algorithm and \textbf{cost(OPT)} be the  measured cost of optimal algorithm where the sequence was known beforehand. In the worst case scenario, first customer arrives at the middle of the connecting lines of two facilities. The adversary sets the next customers on top of the facilities which are recently filled up. 
\begin{figure}[H]
\begin{center}
\includegraphics[width=4cm]{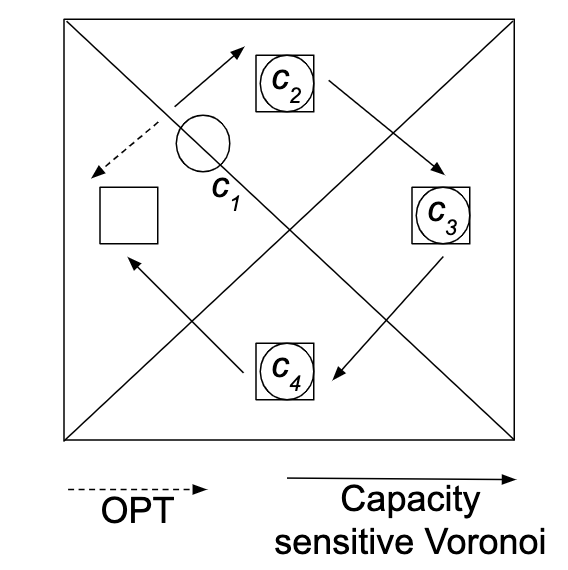}
\end{center}
\caption{Facility Assignment on a Plane}
\label{figure:facility_on_plane}
\end{figure}
Suppose that there are four customers on a plane just like in Figure \ref{figure:facility_on_plane}. Here the first customer $c_1$ will be assigned to the facility to the right. Then the following customer $c_2$ will appear on the facility where the first customer is assigned. the second customer will be assigned to the facility to the right as it is in the figure.
The total cost of assigning four customers will be
$$cost(ALG) = \frac{d}{3\sqrt{2}}+\frac{\sqrt{2}d}{3}+\frac{\sqrt{2}d}{3}+\frac{\sqrt{2}d}{3} = \frac{7}{3\sqrt{2}}d.$$

In this case optimal algorithm would place the first customer to the facility in the left and the total cost would be

$$cost(ALG) = \frac{d}{3\sqrt{2}}+0+0+0 = \frac{1}{3\sqrt{2}}d.$$

Then the competitive ratio would be
$$\mathcal{R}(\text{Algorithm Capacity sensitive voronoi}) = \frac{cost(ALG)}{cost(OPT)} = \frac{7d/3\sqrt{2}}{d/3\sqrt{2}} = 7.$$

We can generalize this approach for $n$ facilities with unit capacity. Suppose there are $n$ facilities with unit capacity which are equally distributed inside the polygon and $n \geq 3$. Let $d$ be the radius of the side length of the polygon and the distance between two adjacent facility is $p$. It is trivial to show that $p = f(d)$. \newline
If the first customer arrives strictly interior to a voronoi edge that does not belong to the convex hull. Our algorithm assigns it to the closest facility, then the next customer arrives on the top of the facility which is occupied by the first customer. So the customer is assigned to the facility right-adjacent to the position of the customer. This process goes on until all the customers arrive and get assigned. The cost of the algorithm is 
$$cost(ALG) = \frac{p}{2}+ \sum_{i=1}^{n-1}p = \frac{2n-1}{2}p.$$
But if the sequence is known to us beforehand, just like an offline algorithm , the optimal assignment would be to choose the alternate facility for the first customer and other customers will have zero cost when assigning the them to the facilities. The cost of OPT would be
$$cost(OPT) =  \frac{p}{2} + 0+0+\dots = \frac{p}{2}.$$

Then the competitive ratio would be
$$\mathcal{R}(\text{Algorithm Capacity sensitive voronoi}) = \frac{cost(ALG)}{cost(OPT)} = \frac{(2n-1)p/2}{p/2} = 2n-1.$$

We now assume the input sequence is not well distributed. The argument is similar to \ref{theorem:greedy_grid}. Whether the spanning area is small or all customers are placed in the same location, the cost does not improve in any case.

\end{proof}

\section{Hardness result} \label{sec:five}
The online facility assignment problem is related to the cow path problem where a cow is searching for the bridge in order to cross a river. We can represent the river by a straight line, where the initial position of the cow is at the center of the line. With out loss of generality we can assume that the cow starts to move $d_1$ steps left from the origin, turn right and moves $d_2$ steps right to the origin, turn again to left and moves $d_3$ steps from the origin ($d_3>d_1$). The cow continues this process until it finds the bridge. We say a cow path is $c$-competitive if the summation of the total steps to left and right is no more than $c$-times than the minimum number of steps.

We can simulate the cow path problem in an instance of thie online facility assignment problem by generating the input sequence of customers in a special way. We consider an instance where the capacity of each facility is equal to one. Hence, only one customer can be assigned to each facility. We assume that the facilities are located on the integer points of the straight line and the customers appear on the integer points in an online manner. Suppose ALG is a $c$-competitive algorithm for the assignment problem. When a customer $c_i$ appears, it is assigned to either the closest free facility $f_l$ at left of $c_i$ or to the closest free facility $f_r$ at right of $c_i$ by algorithm ALG. Note that, if we consider a facility $f'_l$ which is further left of $c_i$ compared to $f_l$ and later another customer $c_j$ is assigned to $f_l$, then the overall assignment cost is never going to decrease. Hence, any algorithm that is trying to minimize the total assignment cost may only consider the closest left and right free facility only.

Now consider the first two customers are placed on the middle facility $f_m$ (center of the straight line). With out loss of generality, we assume that the second customer is assigned to $f_{m-1}$ by ALG. Now the third customer $c_3$ is placed exactly on $f_{m-1}$ and suppose ALG assigns it to $f_{m-2}$. The adversary places the next customer $c_4$ exactly on $f_{m-2}$. This process continues and at some point ALG has to assign the new customer to $f_{m+1}$. Otherwise ALG can not be a $c$-competitive algorithm and later at some point ALG has to switch to left again. Hence, the assignment direction switches left and right similar to a cow path. We show the relationship between an assignment and cow path is the following lemma.
\begin{lemma}
\label{theorem:ofa_cow_path}
Any $c$-competitive algorithm for online facility assignment on a line yields a $c$-competitive algorithm for the cow path problem.
\end{lemma}

The proof is deferred to Appendix~\ref{apdx:ofa_cow_path}. If we set the capacity of each facility equal to one, then the problem is called the online matching problem. Hence, the online facility assignment problem is a generalized version of the online matching problem. Due to the relationship between the cow path problem and the online matching problem it has been conjectured that there exists a $9$-competitive algorithm for the online matching problem similar to the cow path problem~\cite{Kalyanasundaram1998}. However, Fuchs et al. ~\cite{FUCHS2005251} have shown that no algorithm for the online matching problem has competitive ratio less than $9.001$, which immediately yields the following corollary.
\begin{corollary}
\label{theorem:hardness}
Let ALG be an algorithm for online facility assignment on a line. Then $\mathcal{R}(ALG) \geq 9.001$.
\end{corollary}

\section{Conclusion}
We first computed that Algorithm Greedy on grid graph has the competitive ratio of $r \times c + r + c$. We also showed that the competitive ratio of Algorithm Optimal-Fill on grid graph is $O(r \times c)$.
The competitive ratio of Algorithm Optimal-Fill for arbitrary connected unweighted graph is $2|F|$ which is tight and better than the previous result $\frac{|E||F|}{r}$, where $E$ is the set of edge of graph $G$, $F$ is the set of facilities and $r$ is the radius of the graph. \par

Then we provide an algorithm to compute assignment on a plane that has competitive ratio of $(2n-1)$. Finally, we show that no algorithm which assigns facilities on a line has competitive ratio less than $9.001$. The algorithm we studied in this paper has a linear competitive ratio in terms of the number of vertices or facilities. The development of algorithms that has a sublinear or constant competitive ratio remains an interesting open problem.\par

\bibliography{facilityLocation}
\appendix

\section{Proof of Lemma~\ref{lemma:opt_fill_max_vertices}}\label{apdx:opt_fill_max_vertices}
\begin{proof}
Every row can have at most two vertices that have equal distance from $r$. The top and bottom row can have at most one vertex. Hence, the total number of vertices in $V'$ can be at most $2n-2$.
\end{proof}

\section{Proof of Lemma~\ref{lemma:equal_distance_from_center}}\label{apdx:equal_distance_from_center}
\begin{proof}
The claim is true for $r$ equal to $0$ and $1$. Now let the claim be true for $r=k$. We need to show that the claim is true for $r=k+1$. Let $V_k$ be the set of vertices for $r=k$. Hence, $|V_k|=4k$ by induction hypothesis. We now show that $V_{k+1}=4(k+1)$. We consider every row of $V_k$. If a row has one vertex $v$, then the left and right neighbors of $v$ belong to $V_{k+1}$. If a row has two vertices, then there are two vertices in $V_{k+1}$ from that row: one is the right neighbor of the right vertex and another is the left neighbor of the left vertex. Note that, here we assumed that such neighbors exist, otherwise there is no vertex from that row to $V_{k+1}$. There are two rows for $V_k$ having only one vertex. Also, at most two new rows can be considered for $V_{k+1}$ each having one vertex. Hence, we have at most four more vertices in $V_{k+1}$. 
\end{proof}

\section{Proof of Lemma~\ref{lemma:optimal_fill_total_cost}}\label{apdx:optimal_fill_total_cost}
\begin{proof}
We consider the worst case scenario: there is a facility with unit capacity on every vertex of $V'$. The adversary places a customer on the center vertex $c$, Algorithm Optimal-Fill assigns it to the facility situated on $v_1$, see Figure~\ref{figure:opt_fill_r_grid}. The adversary places the next customer on $v_1$, Algorithm Optimal-Fill assigns that customer to the facility on $v_2$. The adversary continues this process, placing new customer to a facility which is already assigned to another customer. Note that, the scenario shown in Figure~\ref{figure:opt_fill_r_grid} is not unique, however in each step Algorithm Optimal fill pays $O(r)$ cost. By Lemma~\ref{lemma:equal_distance_from_center}, there are $O(r)$ such facility in $V'$. Hence, the total assignment cost of Algorithm Optimal-Fill is $O(r^2)$.
\begin{figure}[h]
\begin{center}
\includegraphics[width=4cm]{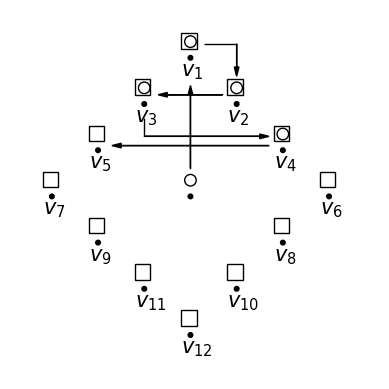}
\end{center}\caption{A costly configurations of Algorithm Optimal-Fill}
\label{figure:opt_fill_r_grid}
\end{figure}
\end{proof}

\section{Proof of Lemma~\ref{lemma:optimal_lower_bound}}\label{apdx:optimal_lower_bound}
\begin{proof}
If each customer is assigned very close to a free facility then optimal assignment and Algorithm Optimal-Fill are the same. In order to force Algorithm Optimal-Fill to pay significantly more compared to an optimal assignment, the adversary has to force Algorithm Optimal-Fill to make large assignments in every step. According to our claim if an assignment of a particular customer by Algorithm Optimal-Fill is large, then the total optimal assignment is at least approximately half of that large assignment cost. We have pictorially shown why this happens in Figure~\ref{fig:opt_fill_basis.pdf}.

\begin{figure}[H]
\begin{center}
\includegraphics[width=12.5cm]{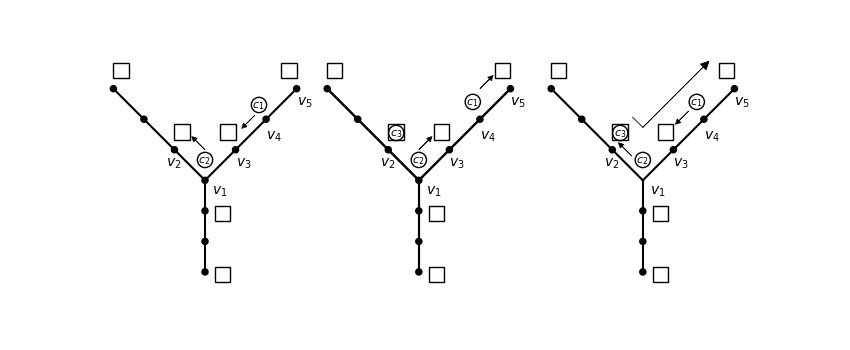}
\end{center}
 \minipage{0.32\textwidth}
  \subcaption{Initially, OPT and Algorithm Optimal-Fill has the same assignment}\label{fig:optimal_fill_a}
\endminipage\hfill
\minipage{0.32\textwidth}
  \subcaption{The optimal assignment changes after the third customer appear}\label{fig:optimal_fill_b}
\endminipage\hfill
\minipage{0.32\textwidth}%
  \subcaption{The assignment of Algorithm Optimal-Fill after the third customer appears}\label{fig:optimal_fill_c}
 \endminipage

\caption{Algorithm Optimal-Fill has a large assignment cost for the third customer that appeared on vertex $v_2$. In Figure (a) we can see that we have two customers on vertex $v_1$ and $v_4$. The assignment costs of both customers for Algorithm Optima-Fill and the optimal algorithm are the same. Both of the assignment costs are relatively small. In Figure (b) a new customer arrives on vertex $v_2$. As the optimal assignments change for both previous customers, Algorithm Optimal-Fill has to pay a large cost for $c_3$. In Figure (c) we see that Algorithm Optimal-Fill assigns the third customer on $v_2$ to the facility on $v_5$ since the optimal algorithm uses this facility in one assignment.}
\label{fig:opt_fill_basis.pdf}
\end{figure}

Lets now explain that formally. In the initial assignment, the adversary can not do very much since all the facilities are free. Suppose the adversary places the first customer $c_1$ around the midpoint of two facilities. For example, $\epsilon_1$ away from the midpoint. Let the midpoint be vertex $v_1$, see Figure~\ref{figure:optimal_fill_eps_a}. Suppose the two closest facility of $c_1$ are $f_1$ and $f_2$. If Algorithm Optimal-Fill assigns $c_1$ to $f_1$ then it will save $2 \epsilon_1$ compared to the assignment cost of $c_1$ to $f_2$. Obviously, Algorithm Optimal-Fill will assign $c_1$ to $f_1$ since it is free. However, we compared the two assignment cost because it will play an important role in the analysis.

Now the adversary will place the second customer $c_2$ also very close to $f_1$. Suppose the two adjacent facilities of $c_2$ are $f_1$ and $f_3$. The distance of $c_2$ from the midpoint of $f_1$ and $f_3$ is $\epsilon_2$. Let the midpoint be vertex $v_2$, see Figure~\ref{figure:optimal_fill_eps_a}. Now the adversary will try to make the assignment cost of $c_2$ as large as possible. One possibility is to assign $c_2$ to $f_3$ since $f_1$ is not free. But the assignment cost of Algorithm Optimal-Fill will be significantly larger if Algorithm Optimal-Fill assigns $c_2$ to $f_2$. In order to make it happen, the adversary has to adjust the optimal assignment since, opt-fill makes its decision based on the optimal assignment. The adversary has to make $\epsilon_2>\epsilon_1$ to force optimal algorithm to assign $c_2$ to $f_1$ and $c_1$ to $f_2$, see Figure~\ref{figure:optimal_fill_eps_a}. Note that, $f_1$ is closer to $c_1$ compared to $f_2$ but still in the optimal assignment $c_1$ is assigned to $f_2$ because the overall assignment cost will decrease. If the assignment cost of $c_2$ by Algorithm Optimal-Fill is $x$, then the total optimal assignment cost is around $x/2$. More specifically, let the distance between $f_1$ and $f_2$ is $x'$. Here, $x$ is a little larger than $x'$ and the cost of assigning $c_1$ by optimal algorithm is larger than $x'/2$. Note that the argument $x$ is a little larger than $x'$ might not be quite reasonable for only two customers, but in a long run the value of $x$ will be dominated by $x'$ where $x'$ is the distance between the facility assigned for the last customer and the closest facility along that direction to the last customer.

Now we transfer this scenario to an equivalent scenario by adjusting the value of $\epsilon_1$ and $\epsilon_2$. We assign $\epsilon_1'=0$ and $\epsilon_2'=\epsilon_2-\epsilon_1$. Note that, the last assignment cost by Algorithm Optimal-Fill and the total assignment cost of optimal assignment remains the same. However, it is more straightforward to argue that the total optimal assignment is half of the last assignment cost of opt-fill. Since, $c_1$ is exactly at the midpoint of $f_1$ and $f_2$. For this simple scenario, this transformation may not be necessary, but we will find it useful later in complicated scenarios, see Figure~\ref{figure:optimal_fill_eps_b}.


\begin{figure}[H]
\minipage{0.48\textwidth}
  \includegraphics[width=\linewidth]{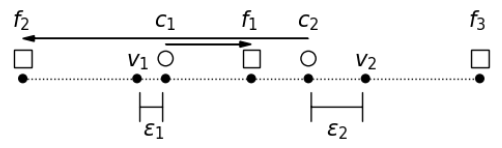}
  \subcaption{Original configuration with two customers.}
\label{figure:optimal_fill_eps_a}
\endminipage\hfill
\minipage{0.48\textwidth}
  \includegraphics[width=\linewidth]{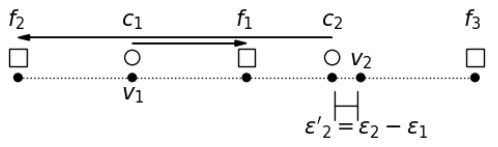}
  \subcaption{An equivalent configuration.}
\label{figure:optimal_fill_eps_b}
\endminipage
\caption{(a) The adversary placing customers around the midpoint of two facilities to harm Algorithm Optimal-Fill. Here, the dotted line indicates that we are drawing a portion of the unweighted graph, there are other vertices in this graph that do not play a significant role to illustrate the idea and not shown in the figure. (b) This configuration is equivalent to Figure~\ref{figure:optimal_fill_eps_a}.}
\end{figure}

Now let us consider that we have more than two customers. Let the current customer is $c_i$. Now suppose $c_i$ is assigned to a facility that is far away from it due to the trick adversary plays as described above. Now let us assume that the facility $c_i$ is assigned by Algorithm Optimal-Fill is $f_i$ and the closest facility in that direction is $f_i'$. Now, suppose the second closest facility to $c_i$ is $f_i''$. Suppose the distance from the midpoint of $f_i'$ and $f_i''$ is $\epsilon_i$. Now, we adjust the values $\epsilon_1, \epsilon_2, \cdots, \epsilon_i$ as described above so that $\epsilon_i'\geq 0$ and $\epsilon_j'=0, j<i$. Now if the distance between $f_i$ and $f_i'$ is $x$, then the total optimal assignment is $x/2$, see Figure~\ref{fig:opt_fill_complex}. Also, the assignment cost of $c_i$ by Algorithm Optimal-Fill is approximately $x$. Hence, the claim is true. 

\begin{figure}[H]
\minipage{0.48\textwidth}
  \includegraphics[width=\linewidth]{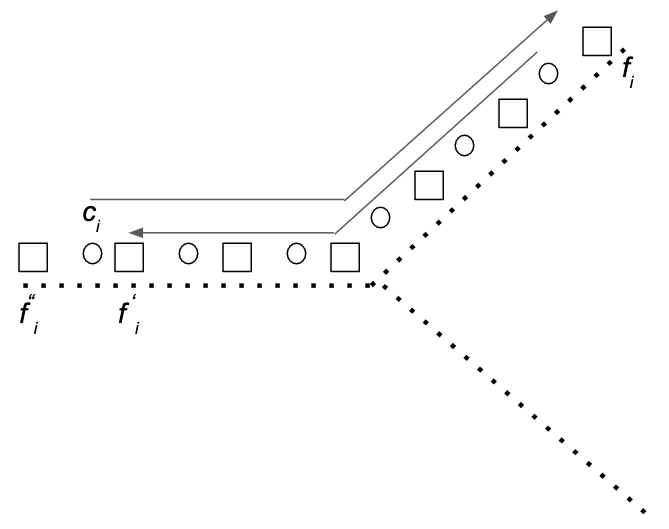}
  \subcaption{The assignment of Algorithm Optimal-Fill.}
\label{figure:optimal_fill_eps_c}
\endminipage\hfill
\minipage{0.48\textwidth}
  \includegraphics[width=\linewidth]{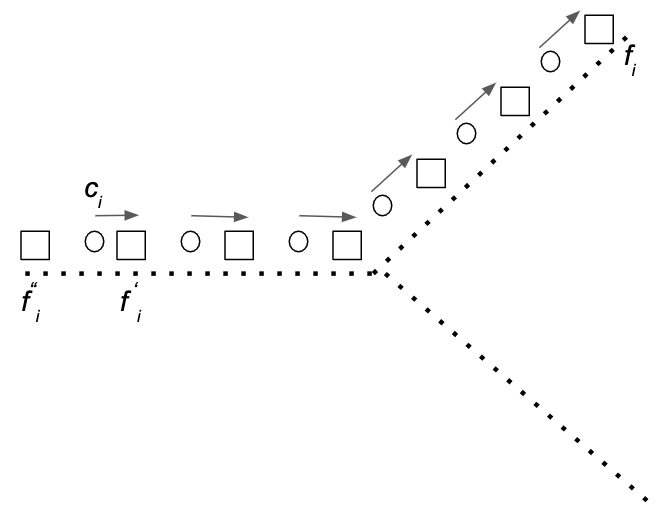}
  \subcaption{The assignment of the optimal algorithm.}
\label{figure:optimal_fill_eps_d}
\endminipage
\caption{(a) The customer $c_i$ is assigned to $f_i$. It can not be assigned to $f_i'$ since another customer has been already assigned to that facility. Note that we do not show the assignments of the remaining customers to keep the figures simple. The customer $c_i$ could be assigned to $f_i''$, but the reason $c_i$ is eventually assigned to $f_i$ is the optimal assignment, see Figure (b).}
\label{fig:opt_fill_complex}
\end{figure}
\end{proof}

\section{Algorithm Capacity sensitive Voronoi diagram}\label{apdx:Voronoi}

\begin{algorithm}[H]
\caption{Capacity sensitive Voronoi diagram} \label{Voronoi}
\begin{algorithmic}
 \State $n \gets $ number of facilities 
 \State  $F[ ] \gets $ array of facility
 \State $C[ ] \gets $ array of the capacity of the facilities
 \State  $cost \gets $ 0
 
 \For{$i\gets 1$ to $n$}
     \State $F[i] \gets $ co-ordinate of the $i$-th facility 
     \State $C[i] \gets $ capacity of the $i$-th facility 
 \EndFor

 \While{customers are coming}
     \State $locate \gets$ co-ordinate of the location of customer
      \State Draw a weighted Voronoi diagram using $F$ and $C$
     \State $k \gets $ index of the facility's area of influence 
     \State $cost \gets  cost+distance(F[k], locate)$ 
     \State $C[k] \gets C[k]-1$ 
     \If{$ C[k] == 0$}
         \State remove the facility $F[k]$ from $F$
     \EndIf
 \EndWhile
\end{algorithmic}
\end{algorithm}

\section{Proof of Lemma~\ref{theorem:ofa_cow_path}}\label{apdx:ofa_cow_path}
\begin{proof}
The adversary starts by placing two customers exactly on the middle facility. The remaining customers are placed exactly on the facilities that has been used for the immediate previous customers. The optimal assignment cost is distance between the middle facility and the last facility. In the instance of the cow path problem the bridge is situated at the location of the last facility assigned. In both instances the total cost of the algorithm and the optimal cost is same. Hence, an algorithm for the assignment problem generates a valid cow path having same competitive ratio.
\end{proof}

\end{document}